\documentclass[letterpaper,11pt]{article}

\usepackage[utf8]{inputenc} 
\usepackage{fullpage,physics,graphicx,xcolor,bm,bbm,
authblk,mathtools,amsmath,amsfonts,amsbsy,amssymb,hyperref}
\usepackage[british]{babel}
\usepackage[title]{appendix}
\usepackage[square,numbers,sort&compress]{natbib}
\hypersetup{colorlinks = true, linkbordercolor = white, linkcolor = blue, citecolor = blue}
\newcommand{\equa}[1]{Eq.~(\ref{#1})} \newcommand{\equas}[1]{Eqs.~(\ref{#1})}
\newcommand{\equass}[2]{Eqs.~(\ref{#1})-(\ref{#2})}
\newcommand{\equasa}[2]{Eqs.~(\ref{#1}) and (\ref{#2})}
  
 \newcommand{\Ra}{{\rm Ra}}  \newcommand{\Fr}{{\rm Fr}}
\newcommand{\eqn}[2]{\begin{gather}
#1
\label{#2}
\end{gather}
}
\newcommand{\gat}[2]{\begin{subequations}\label{#2}\begin{gather}
#1
\end{gather}\end{subequations}
}

\title{\bf On the use and misuse of the Oberbeck--Boussinesq approximation}
\author{\bf A. Barletta$^1$\footnote{Corresponding author: \texttt{antonio.barletta@unibo.it}}\ \ 
; M. Celli$^1$\ ; D.A.S. Rees$^2$}
\affil{\small 
$^1$\ Department of Industrial Engineering, Alma Mater Studiorum Universit\`a di Bologna,\\
Viale Risorgimento 2, 40136 Bologna, Italy.\\
\vspace{1.5mm}
$^2$\ Department of Mechanical Engineering, University of Bath, Bath BA2 7AY, UK.
}
\date{\small\today} 

\begin{document}

\maketitle

\begin{abstract}\noindent
The Oberbeck--Boussinesq approximation is the most widely employed theoretical scheme for the study of natural or mixed convection flows. However, the misunderstanding of this approximated framework is a possibility that may cause the emergence of paradoxes or, at least, incorrect conclusions. In this note, the basic features of the Oberbeck--Boussinesq approximation are  briefly recalled and three simple examples where this theoretical scheme may be misused are provided. Such misuses of the approximation lead to erroneous conclusions that, in the examples presented in this note, entail violations of the principle of mass conservation. A discussion about the Oberbeck--Boussinesq approximation as an asymptotic theory obtained by letting the product of the thermal expansion coefficient and the reference temperature difference tend to zero is also presented.

~

\end{abstract}

\section{Introduction}
The study of natural or mixed convection flows either in fluids or in fluid--saturated porous media is, with a limited number of exceptions, modelled theoretically by claiming the validity of the Oberbeck--Boussinesq approximation. There are several thorough and comprehensive analyses of how this approximation can be established starting from a general formulation of the local balance equations of mass, momentum and energy for a fluid. Beyond the many textbooks of fluid dynamics and convection heat transfer, we mention the analyses of this topic presented in chapter 8 of \citet{zeytounian1990asymptotic}, in \citet{rajagopal1996oberbeck} and in \citet{ZEYTOUNIAN2003575}. Such discussions on the origin and on the range of applicability of the Oberbeck--Boussinesq approximation stem from pioneering papers such as \citet{spiegel1960boussinesq}, \citet{gray1976validity} and  \citet{hillsroberts1991}. A recent interesting review on the topic has been presented by \citet{mayeli2021buoyancy}.

The aim of this short paper is to highlight an aspect of the Oberbeck--Boussinesq approximation that may be the source of pitfalls, i.e., the duality of variable fluid density and constant fluid density. If it is recognised that the approximation scheme predicates a variable density which is pressure independent and linearly varying with the temperature, in some instances one may forget that such variable density serves only to define the buoyancy force within the local momentum balance equation. Utilising the variable density outside this very specific context may lead to unphysical predictions and, hence, to incorrect conclusions. Such conclusions are incorrect as they usually conflict with the principle of mass conservation. A final discussion about the interpretation of the Oberbeck--Boussinesq approximation as a limiting case of the general local balance equations of mass, momentum and energy is presented.

\section{A minimalistic survey of the Oberbeck--Boussinesq model}
There are several detailed and thorough descriptions of the Oberbeck--Boussinesq model for natural and mixed convection flows. The basic system of partial differential equations, expressing the local mass, momentum and energy balances, is given by
\gat{
\div{\vb{u}} = 0, \label{1a}\\
\rho_0 \qty[\pdv{\vb{u}}{t} + \qty(\vb{u} \vdot \grad) \vb{u}] = - \grad{P} + \rho\, \vb{g} + \mu\, \laplacian \vb{u}, \label{1b}\\
\pdv{T}{t} + \qty(\vb{u} \vdot \grad) T = \alpha\, \laplacian T , \label{1c}
}{1}
where $\vb{u}$ is the velocity, $P$ is the pressure, $T$ is the temperature, $t$ is time, $\mu$ is the dynamic viscosity and $\alpha$ is the thermal diffusivity. There are two densities in \equas{1}. One is the reference density $\rho_0$, i.e., the fluid density evaluated at the reference temperature $T_0$. On the other hand, $\rho$ denotes the fluid density evaluated at the local temperature $T$ through the linear equation of state,
\eqn{
\rho = \rho_0\, \qty[ 1 - \beta\, \qty(T - T_0) ],
}{2}
where $\beta$ is the thermal expansion coefficient. We mention that the linear equation of state (\ref{2}) is to be replaced by a quadratic equation of state in special situations such as pure water close to $4^{\circ}{\rm C}$.

It must be stressed that, in \equasa{1}{2}, $\qty(\rho_0, \mu, \alpha, \beta)$ are constant, pressure-independent, fluid properties evaluated at the reference temperature $T_0$. In order to maximise the reliability of the approximation, a judicious choice of the reference temperature is an average temperature over the flow spatial domain and over the time interval of the flow process. We also implicitly assume, for the local energy balance (\ref{1c}), that the viscous heating effect is negligible and that no internal heat source is present.

When the Oberbeck--Boussinesq scheme is to be applied to the seepage flow in a porous medium, then the local momentum balance equation may be modelled through Darcy's law, namely
\eqn{
\frac{\mu}{K}\, \vb{u} = - \grad{P} + \rho\, \vb{g},
}{3}
where $\vb{u}$ now denotes the seepage, or Darcy's, velocity and $K$ is the permeability of the medium. Then, \equa{3} supersedes \equa{1b}, in this case. Also \equa{1c} is to be reformulated when we study the seepage flow in a porous medium. In fact, the local energy balance reads
\eqn{
\sigma\,\pdv{T}{t} + \qty(\vb{u} \vdot \grad) T = \alpha_m \laplacian T , 
}{3n}
where $\sigma$ is the ratio between the average heat capacity of the saturated porous medium and the heat capacity of the fluid, while $\alpha_m$ is the average thermal diffusivity of the saturated porous medium.

\subsection{Hydrostatic pressure, buoyancy force}
The gravitational force term, $\rho \vb{g}$ in either \equa{1b} or in \equa{3}, is the one and only quantity where the use of \equa{2} is allowed. Thus, we can write
\eqn{
- \grad P + \rho\, \vb{g} = - \grad\Big[P - \underbrace{\qty(P_e + \rho_0\, \vb{g} \vdot \vb{x})}_{\displaystyle{P_h}}\Big] + \vb{b} ,\nonumber\\[6pt]
- \grad P + \rho\, \vb{g} = - \grad\qty(P - P_h) + \vb{b} ,
}{4} 
where
\eqn{
P_h = P_e + \rho_0\, \vb{g} \vdot \vb{x}
}{5}
is the hydrostatic pressure, with $\vb{x} = \qty(x,y,z)$ denoting the position vector, and
\eqn{
\vb{b} = - \rho_0\, \beta\, \qty(T - T_0)\, \vb{g}
}{6}
is the buoyancy force. Both $P$ and $P_h$ depend on the spatial position. It is a common practice to use a specific notation for their local difference,
\eqn{
p = P - P_h .
}{7}
We call $p$ hydrodynamic pressure, for the sake of brevity, to mark the distinction from the pressure and the hydrostatic pressure. In \equa{5}, $P_e$ is an arbitrary constant value which, physically, can be assigned in order to fix a possible overall external pressurisation of the fluid. In most usual cases, $P_e$ is assumed to coincide with the atmospheric pressure. 

Out of the present context, where the hydrostatic pressure and the buoyancy force are defined on the basis of a chosen reference temperature, $T_0$, the fluid density is considered to be a constant equal to $\rho_0$. If one violates this simple rule, then erroneous conclusions could be drawn which will typically lead to violations of the mass conservation principle.

\section{A rectangular cavity with side heating}
\begin{figure}[t]
\centering
\includegraphics[width=0.6\textwidth]{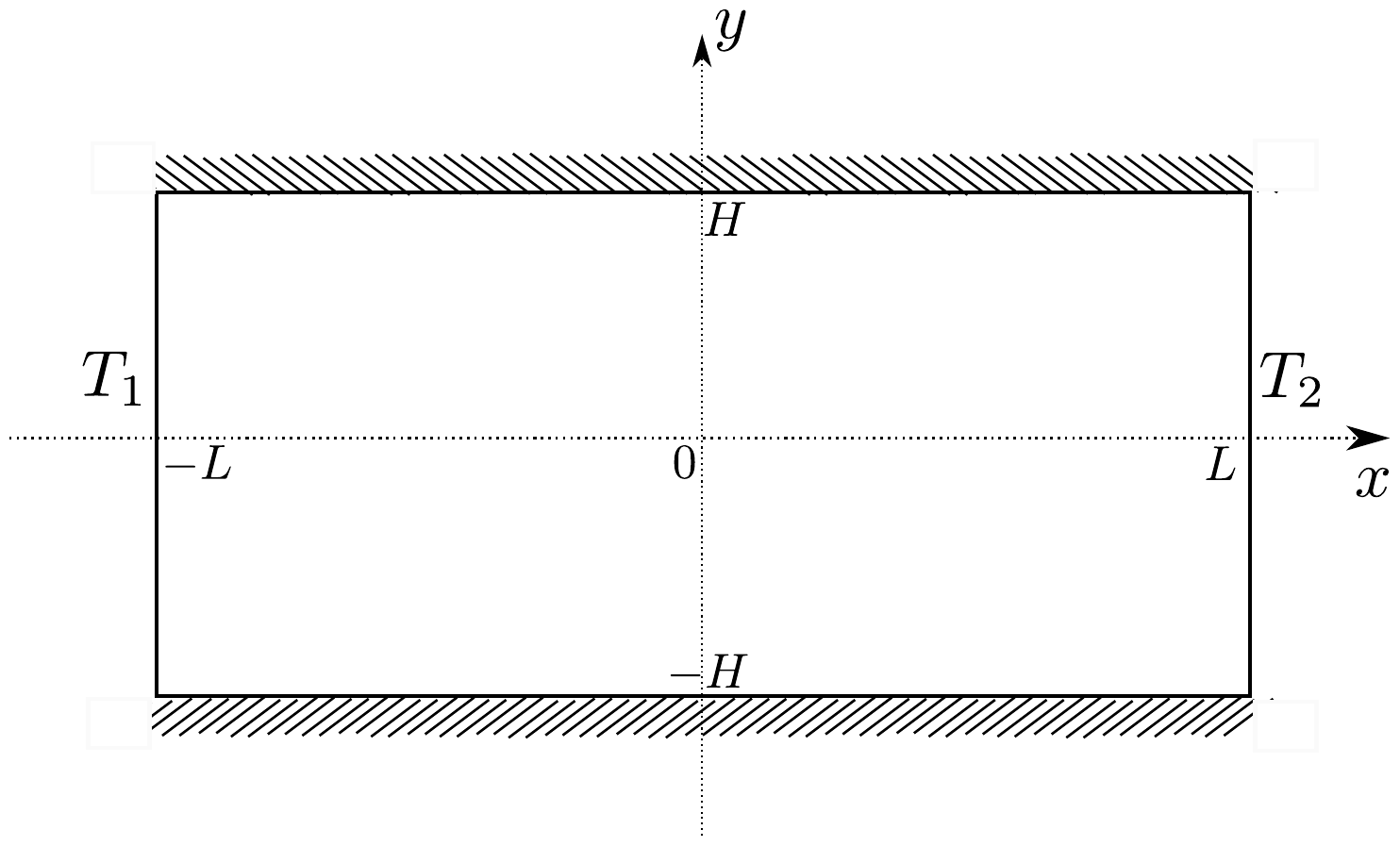}
\caption{\label{fig1}Sketch of the rectangular cavity with side heating.}
\end{figure}

Let us consider the classical problem of two--dimensional natural convection in a rectangular cavity with rigid and impermeable side boundaries kept at uniform, but different, temperatures, $T_1$ and $T_2$, while the upper and lower sides are kept adiabatic. It is not restrictive to assume $T_1 > T_2$. A sketch of the system is provided in Fig.~\ref{fig1}.  

It is well--known that, with sufficiently small differences $T_1 - T_2$, a steady--state natural convection flow occurs in the cavity. In non--dimensional terms, this restriction is equivalent to a sufficiently small Rayleigh number. The steady--state flow is cellular in character, with one or more convective cells. For a given fluid, the number of stationary convective cells depends on the aspect ratio, $H/L$, and on the Rayleigh number. If we denote with $S$ the horizontal midplane surface with $y=0$ and $x \in \qty[-L, L]$, then we can immediately conclude that
\eqn{
\int\limits_S v \, \dd S = 0 \qor \int\limits_{-L}^L v \, \dd x = 0,
}{8}
as an obvious consequence of \equa{1a}. In fact, one can just integrate \equa{1a} over the upper half--domain $\{x \in [-L,L], y \in [0,H]\}$ and then employ Gauss' theorem by recalling that $S$ is the only permeable boundary of such a domain. Here, $v$ is the $y$ component of $\vb{u}$. Then, we could wonder how we could evaluate the mass flow rate, $\dot m$, per unit depth (in the $z$ direction) across $S$. There is a correct way and an incorrect way. The correct way is by applying the principle that the fluid density is to be considered constant and equal to $\rho_0$, so that \equa{8} yields
\eqn{
\dot{m} = \rho_0 \int\limits_{-L}^L v \, \dd x = 0.
}{9}
The incorrect way is by employing the variable density expressed by \equa{2}, so that \equa{8} yields
\eqn{
\dot{m} = \rho_0 \int\limits_{-L}^L v \, \dd x - \rho_0\, \beta \int\limits_{-L}^L \qty( T - T_0)\, v \, \dd x = - \rho_0\, \beta \int\limits_{-L}^L \qty( T - T_0)\, v \, \dd x < 0.
}{10}
The reason why \equa{10} yields $\dot m < 0$ is that, at least with a sufficiently small temperature difference $T_1 - T_2$, both $v$ and $T - T_0$ are odd functions of $x$ with $x \in \qty[-L, L]$, positive for $x \in (-L,0)$ and negative for $x \in (0,L)$ (see, for instance, \citet{de1968laminar}). 
Thus, their product is an even and positive function of $x$ throughout the domain of integration $x \in \qty[-L, L]$, so that \equa{10} leads to the conclusion that $\dot m < 0$. 
The numerical solution discussed by \citet{de1968laminar} is complemented by a simple analytical solution which predicts the same symmetry for $v$ and $T-T_0$ and which holds for a very tall cavity, $H \gg L$. Such an asymptotic solution is briefly outlined in Appendix~\ref{Appe}. We mention that \equa{8} holds also for any other $y = constant$ plane $S$, as a consequence of \equa{1a}. 

We said that \equa{10} expresses the incorrect way to evaluate $\dot m$ since the conclusion $\dot m \ne 0$ is an evident violation of the mass conservation within the upper half--domain $\{x \in [-L,L], y \in [0,H]\}$ or in the lower half--domain $\{x \in [-L,L], y \in [-H,0]\}$. In fact, both for the upper and the lower half--domains, $S$ would be the only permeable boundary and it would be crossed by a net mass flow rate. Such a situation, in a stationary regime, yields a violation of the principle of mass conservation. 

\begin{figure}[t]
\centering
\includegraphics[width=0.5\textwidth]{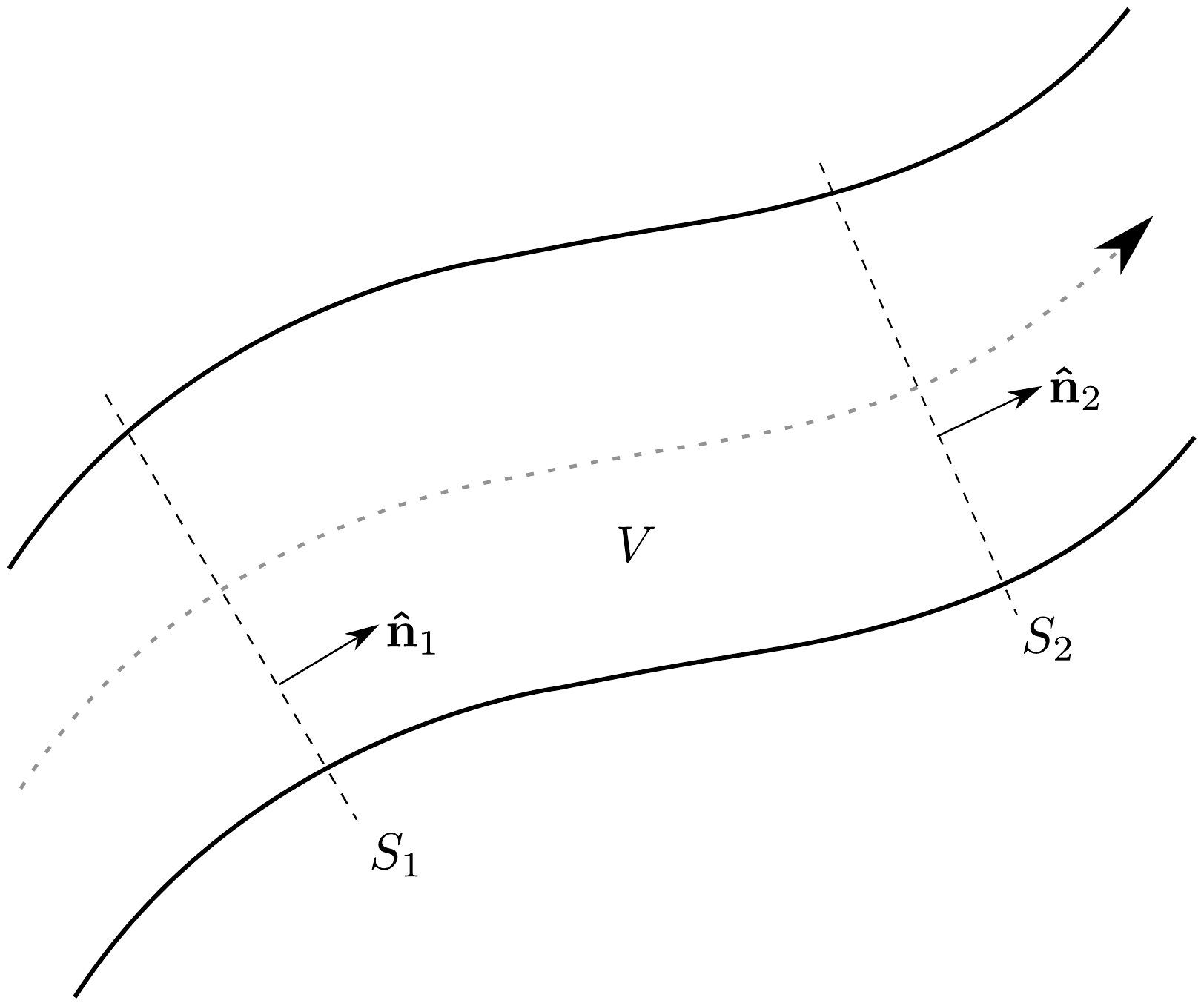}
\caption{\label{fig2}Sketch of mixed convection duct flow.}
\end{figure}

\section{Mixed convection duct flow}
Let us consider the internal mixed convection in a duct with an impermeable wall having an increasing temperature along the streamwise direction. Such a behaviour is observed, for instance, when the duct wall is subject to an incoming uniform heat flux. We consider the case of stationary flow. 

As sketched in Fig.~\ref{fig2}, we consider the region $V$ delimited by the cross--sections $S_1$ and $S_2$. Since, the fluid is heated in the streamwise direction, we have an average temperature at the cross--section $S_1$, denoted by $T_1$, smaller than the average temperature $T_2$ evaluated at the cross--section $S_2$. Then, a judicious choice of the reference temperature $T_0$ for the Oberbeck--Boussinesq approximation in the domain $V$ is the volume--averaged temperature,
\eqn{
T_0 = \int\limits_V T \, \dd V .
}{11}
Such a volume--averaged temperature value, $T_0$, is larger than $T_1$ and smaller than $T_2$. Let us evaluate the average velocities across $S_1$ and $S_2$,
\eqn{
u_{m1} = \frac{1}{S_1} \int\limits_{S_1} \vb{u} \vdot \vu{n}_1 \, \dd S \qc
u_{m2} = \frac{1}{S_2} \int\limits_{S_2} \vb{u} \vdot \vu{n}_2 \, \dd S,
}{12}
where $\vu{n}_1$ and $\vu{n}_2$ are the unit vectors of the surfaces $S_1$ and $S_2$, oriented in the streamwise direction as shown in Fig.~\ref{fig2}. Equation~(\ref{1a}), after an integration over $V$ and use of Gauss' theorem yields the equality $S_1\,u_{m1} = S_2\, u_{m2}$. One can wonder how one can evaluate the mass flow rates across $S_1$ and $S_2$, i.e., $\dot{m}_1$ and $\dot{m}_2$, respectively. The right way is by assuming that the fluid density is to be considered constant and equal to $\rho_0$ all over $V$, so that
\eqn{
\dot{m}_1 = \rho_0 \int\limits_{S_1} \vb{u} \vdot \vu{n}_1 \, \dd S \qc
\dot{m}_2 = \rho_0 \int\limits_{S_2} \vb{u} \vdot \vu{n}_2 \, \dd S.
}{13}
Since $S_1\,u_{m1} = S_2\, u_{m2}$, \equasa{12}{13} allow one to conclude that $\dot{m}_1 = \dot{m}_2$, which is in perfect agreement with the principle of mass conservation.

The incorrect way to evaluate $\dot{m}_1$ and $\dot{m}_2$ is by employing the variable density $\rho$, \equa{2}. In this case, we have
\eqn{
\dot{m}_1 = \rho_0 \int\limits_{S_1} \vb{u} \vdot \vu{n}_1 \, \dd S - \rho_0\, \beta \int\limits_{S_1} \qty(T - T_0)\, \vb{u} \vdot \vu{n}_1 \, \dd S , \nonumber\\
\dot{m}_2 = \rho_0 \int\limits_{S_2} \vb{u} \vdot \vu{n}_2 \, \dd S - \rho_0\, \beta \int\limits_{S_2} \qty(T - T_0)\, \vb{u} \vdot \vu{n}_2 \, \dd S .
}{14}
From \equa{14} and from the equality $S_1\,u_{m1} = S_2\, u_{m2}$, one can write
\eqn{
\dot{m}_1 - \dot{m}_2 = - \rho_0\, \beta \, \qty( ~\int\limits_{S_1} T\, \vb{u} \vdot \vu{n}_1 \, \dd S - \int\limits_{S_2} T\, \vb{u} \vdot \vu{n}_2 \, \dd S) .
}{15}
By employing again the equality $S_1\,u_{m1} = S_2\, u_{m2}$ and by recalling that the average temperature over $S_1$ is smaller than the average temperature over $S_2$, one can immediately conclude that $\dot{m}_1 - \dot{m}_2 > 0$, i.e., the mass rate flowing through $S_1$ is larger than the mass rate flowing through $S_2$ which, in a stationary regime, means a violation of the principle of mass conservation.

\section{A vertical porous slab separating two fluid reservoirs}
\begin{figure}[t]
\centering
\includegraphics[width=0.4\textwidth]{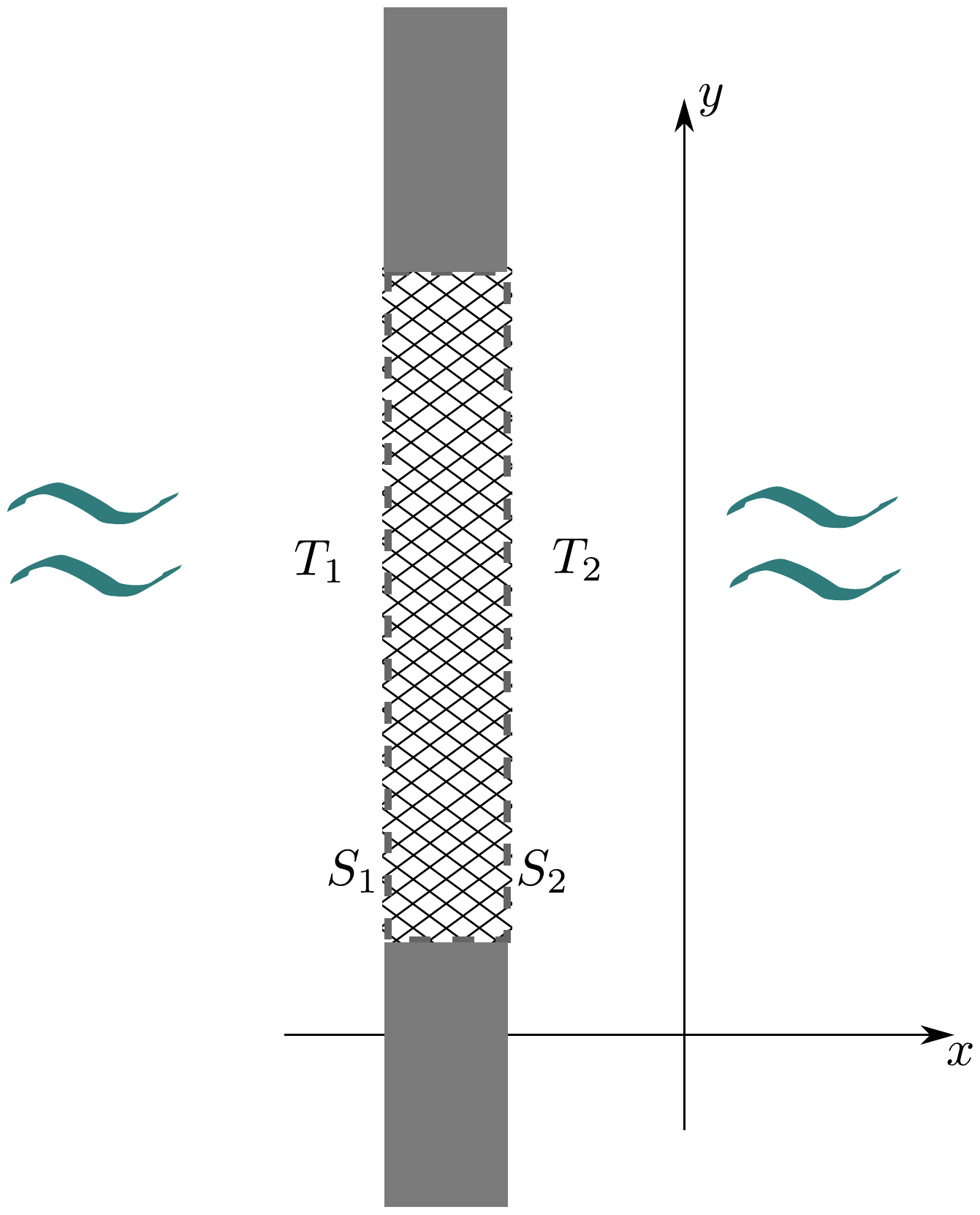}
\caption{\label{fig3}Sketch of of the porous slab separating two fluid reservoirs.}
\end{figure}

Let us consider an infinitely wide wall separating two fluid reservoirs kept at different temperatures, $T_1$ and $T_2$. As shown in Fig.~\ref{fig3}, the wall has a porous insertion bounded by two planes $S_1$ and $S_2$. The fluid in the left--hand reservoir is the same as that in the right--hand reservoir and they are both in a rest state. The same fluid saturates the porous slab. The Oberbeck--Boussinesq approximation can be applied assuming the reference temperature $T_0 = \qty(T_1 + T_2)/2$.

One can evaluate the pressure distribution on the boundaries $S_1$ and $S_2$ as the hydrostatic pressure. By employing \equa{5}, the pressure distribution on the plane $S_1$ is given by
\eqn{
P_1 = P_{e1} - \rho_0 \, g \, y,
}{16}
while, on $S_2$, we have
\eqn{
P_2 = P_{e2} - \rho_0 \, g \, y.
}{17}
Equations~(\ref{16}) and (\ref{17}) have been used to formulate the pressure conditions at the porous slab boundaries by \citet{barletta2015proof, barletta2016instability, barletta2018horton} and by \citet{barletta2019onset}.
The pressurisation constants, $P_{e1}$ and $P_{e2}$, can be equal or not depending on the conditions externally imposed on the two reservoirs. For instance, one may have a situation where the reservoir at temperature $T_1$ is compressed on its top, while the reservoir at temperature $T_2$ is open to the atmospheric pressure. In such a situation, one has $P_{e1} > P_{e2}$. Equations~(\ref{16}) and (\ref{17}) reveal that the pressure distribution on $S_1$ and $S_2$ is $y$ dependent, but the pressure difference across the porous slab is a constant,
\eqn{
\Delta P = P_1 - P_2 = P_{e1} - P_{e2}.
}{18}
If none of the two reservoirs is pressurised, then $\Delta P = 0$. If $\Delta P \ne 0$, one may have a seepage throughflow across the porous slab \cite{barletta2016instability}. 

As sketched in Fig.~\ref{fig3} the porous insertion is bounded above and below by an impermeable material, so that application of \equa{1a} yields
\eqn{
\int\limits_{S_1} u \, \dd S = \int\limits_{S_2} u \, \dd S ,
}{19}
where $u$ is the $x$ component of the seepage velocity, $\vb{u}$, across the porous material. By recalling that the fluid density is to be intended as uniform with the value $\rho_0$, \equa{19} implies that the mass flow rates across $S_1$ and $S_2$ are equal as required by mass conservation,
\eqn{
\dot{m} = \rho_0 \int\limits_{S_1} u \, \dd S = \rho_0 \int\limits_{S_2} u \, \dd S .
}{20}
This is the correct and consistent application of the Oberbeck--Boussinesq scheme to this sample case. 

There is always the possibility of introducing errors as in the previous examples. One may enforce the validity of \equa{2} in the evaluation of the hydrostatic pressures in the two reservoirs so that \equasa{16}{17} are replaced by
\eqn{
P_1 = P_{e1} - \rho_0 \, g \, y \, \qty[1 - \frac{1}{2}\, \beta \, \qty(T_1 - T_2)],
}{21}
and
\eqn{
P_2 = P_{e2} - \rho_0 \, g \, y \, \qty[1 + \frac{1}{2}\, \beta \, \qty(T_1 - T_2)].
}{22}
Equations~(\ref{21}) and (\ref{22}) yield an $y$ dependent pressure difference across the porous layer
\eqn{
\Delta P = P_1 - P_2 = P_{e1} - P_{e2} + \rho_0 \, \beta \, \qty(T_1 - T_2)\, g \, y.
}{23}
Even in the absence of pressurisation in one of two reservoirs $(P_{e1} = P_{e2})$, \equa{23} entails a pressure difference across the porous layer which, in turn, leads to the prediction of a horizontal throughflow across the porous layer \cite{vynnycky_mitchell_2022}. Such a stationary throughflow is unphysical as it would lead to a violation of mass conservation. In fact, consistently with the assumption of a variable density, as given by \equa{2}, the mass flow rate across $S_1$ is expressed as
\eqn{
\dot{m}_1 = \rho_0\, \qty[1 - \frac{1}{2}\, \beta \, \qty(T_1 - T_2)] \int\limits_{S_1} u \, \dd S,
}{24}
while the mass flow rate across $S_2$ is given by
\eqn{
\dot{m}_2 = \rho_0\, \qty[1 + \frac{1}{2}\, \beta \, \qty(T_1 - T_2)] \int\limits_{S_2} u \, \dd S.
}{25}
The violation of mass conservation is quantified by a relative error
\eqn{
\delta = \frac{\dot{m}_2 - \dot{m}_1}{\qty(\dot{m}_2 + \dot{m}_1)/2} = \beta\, \qty(T_1 - T_2) ,
}{26}
where \equas{19}, (\ref{24}) and (\ref{25}) have been employed. Interestingly enough, should $\delta$ be considered as negligible, $\delta \ll 1$, then the pressure distributions given by \equasa{21}{22}, and erroneously employed in the study carried out by \citet{vynnycky_mitchell_2022}, would turn out to match perfectly the correct pressure distributions expressed by \equasa{16}{17}. More precisely, the corrective terms introduced in \equasa{21}{22}, i.e. $\mp\, \delta/2$, can be taken into account consistently only if the solenoidal constraint for the velocity, $\div{\vb{u}}=0$, is replaced by the variable--density local mass balance equation, $\div\qty(\rho\, \vb{u})=0$. The intermediate approximation used by \citet{vynnycky_mitchell_2022}, where \equasa{21}{22} are employed in combination with $\div{\vb{u}}=0$, is flawed as it leads to a violation of the principle of mass conservation.

\section{The Froude number and the Rayleigh number}\label{FrRa}
Let us reconsider \equass{4}{6}. The pressure gradient and gravitational force contributions to the local momentum balance are given by the force per unit volume
\eqn{
\vb{F} = - \grad{P} - \rho_0\, g\, \vu{e}_y + \rho_0\, \beta\, g\, \qty(T - T_0)\, \vu{e}_y ,
}{27}
where we are assuming that the $y$ axis is vertical and orientated upward.
We can rescale $\vb{F}$ in order to obtain a dimensionless formulation of the local momentum balance equation. The dimensionless $\vb{F}$ denoted with an asterisk is given by
\eqn{
\vb{F}^* = \frac{\vb{F}}{\Delta P_r/L} , 
}{28}
where the constant $\Delta P_r$ is a reference pressure difference and the constant $L$ is a reference length. One can also define the dimensionless coordinates pressure and temperature as
\eqn{
(x^*, y^*, z^*) = \frac{(x,y,z)}{L} \qc P^* = \frac{P}{\Delta P_r} \qc T^* = \frac{T - T_0}{\Delta T_r},
}{29}
where $\Delta T_r$ is a reference temperature difference. On account of \equasa{28}{29}, \equa{27} yields
\eqn{
\vb{F}^* = - \grad^*{P^*} - \frac{1}{\Fr}\, \vu{e}_y + \frac{\delta}{\Fr}\,T^* \vu{e}_y ,
}{30}
where the Froude number is defined as
\eqn{
\Fr = \frac{\Delta P_r}{\rho_0\, g\, L},
}{31}
and $\delta$ is given by
\eqn{
\delta = \beta\, \Delta T_r.
}{31bis}
It is to be mentioned that the usual definition is $\Fr = U_0^2/(g\, L)$ (see, for instance, \citet{mayeli2021buoyancy}), where $U_0$ is a reference velocity. In fact, \equa{31} matches perfectly the usual definition provided that one chooses $U_0 = \sqrt{\Delta P_r/\rho_0}$. It is also to be mentioned that, according to other authors (e.g. \citet{zeytounian1990asymptotic}), $\Fr$ should be defined as a velocity ratio, so that it is given by the square root of the quantity identified by \citet{mayeli2021buoyancy} as the Froude number. In our discussion, we will rely on the definition given by \equa{31}.

In the case of natural convection, where buoyancy alone causes the flow, a typical choice of $\Delta P_r$ is
\eqn{
\Delta P_r = \frac{\mu\, \alpha}{L^2} .
}{32}
Thus, the ratio $\delta/\Fr$ coincides with the Rayleigh number,
\eqn{
\Ra = \frac{\delta}{\Fr} = \frac{\rho_0\, g\, \beta\, \Delta T_r\, L^3}{\mu\,\alpha}.
}{33}
In the discussions available in the literature about the rigorous derivation of the Oberbeck--Boussinesq approximation, the approximate governing equations (\ref{1}) are considered as a limiting case of the local balance equations for a fully--compressible (variable density) flow when
\eqn{
\delta \to 0 \qc \Fr \to 0 \quad \text{with} \quad \frac{\delta}{\Fr} \sim \order{1} .
}{34}
This double limit is stated, although in slightly different terms, by  \citet{hillsroberts1991} and reported by \citet{ZEYTOUNIAN19891361, ZEYTOUNIAN2003575}. It is also implicitly employed in the analysis carried out by \citet{rajagopal1996oberbeck}. Also the recent study by \citet{vynnycky_mitchell_2022} relies on this limiting scheme as a basis for the Oberbeck--Boussinesq approximation. However, this approach leads to a singular behaviour of $\vb{F}^*$, as one can easily infer from \equa{30}, since the term $-\qty(1/\Fr)\, \vu{e}_y$  blows up in the limit given by \equa{34}. The singular nature of the Oberbeck--Boussinesq limit of $\vb{F}^*$ is unavoidable, even if it could have been concealed should one have employed, on account of \equasa{5}{7}, the gradient of the dynamic pressure, $\grad p$, instead of the pressure gradient, $\grad P$, on writing the expression of $\vb{F}$ in \equa{27}. However, this trick would have simply swept the dust under the carpet without actually solving the problem. In fact, there is just one way to avoid a singular limiting bahaviour of $\vb{F}^*$ in the limit defined by \equa{34}. One should constrain $\grad{P}$ to be equal to $-\rho_0\, g\, \vu{e}_y$ whenever the Oberbeck--Boussinesq approximation is used. Unfortunately, such a constraint is unphysical except for a quite limited number of special cases.

Interestingly enough, the derivation of the Oberbeck--Boussinesq approximation presented by \citet{rajagopal1996oberbeck} leads to the same unphysical conclusion drawn above, namely that the pressure gradient must always coincide with the hydrostatic pressure gradient. 
We mention that the treatment presented by \citet{rajagopal1996oberbeck} is based on a dimensionless scaling of the governing balance equations where the reference length is $\order{\epsilon^{-1}}$ and the reference velocity is  $\order{\epsilon}$, where $\epsilon$ is the perturbation parameter. The Oberbeck--Boussinesq approximation is defined as the asymptotic case where $\epsilon \to 0$. Thus, the dimensionless scaling is singular in this limit. Incidentally, $\epsilon$ is proportional to $\delta^{1/3}$, with $\delta$ given by \equa{31bis}. The paper by \citet{rajagopal1996oberbeck} discusses in detail the serious drawbacks of the previous theoretical studies that define the Oberbeck--Boussinesq approximation as a limiting case obtained by letting one or more perturbation parameters to zero. Examples are the papers by \citet{spiegel1960boussinesq}, by \citet{gray1976validity} and by \citet{hillsroberts1991}. As a consequence, one can say that an asymptotic theory, based on a suitable perturbation scheme, which is aimed at a rigorous deduction of the Oberbeck--Boussinesq set of governing equations (\ref{1}) is still lacking. It is also possible that the existing approaches that define the Oberbeck--Boussinesq approximation as a limiting case of the fully compressible set of local balance equations are intrinsically biased. It is the authors' opinion that, until a rigorous and nonsingular theoretical scheme will be set up to justify the approximation as an asymptotic regime, its validity relies entirely on the widely--documented experimental validations available in the literature for a very broad range of flow regimes.

\section{Conclusions}
The Oberbeck--Boussinesq approximation for the local balance equations of mass, momentum and energy of buoyancy--induced fluid flows has been briefly outlined. It has been stressed that the linear temperature--dependent density expression is to be used, within the local momentum balance equation, only to transform the combined pressure gradient force and gravity force into a combination of terms involving the dynamic pressure gradient and the buoyancy force. Out of this very specific use, the variable density expression cannot be employed consistently with the approximated Oberbeck--Boussinesq scheme. Three examples have been discussed, relative to either natural or mixed convection flows, where it has been shown that an improper use of the variable density unavoidably leads to violations of the principle of mass conservation. The third of these examples points to a misuse of the Oberbeck--Boussinesq approximation presented in a recent paper \cite{vynnycky_mitchell_2022}. A final section has been devoted to the idea of the Oberbeck--Boussinesq approximation as a limiting case of the fully--compressible local balance equations where the dimensionless parameter $\delta = \beta\, \Delta T_r$ tends to zero, with $\beta$ the thermal expansion coefficient and $\Delta T_r$ the reference temperature difference. It has been pointed out that such an idea leads either to a singular behaviour for the gradient of the dynamic pressure, or to the unphysical constraint that such gradient be identically zero. We conclude that, so far, the Oberbeck--Boussinesq approximation is to be classified as a phenomenologically based model.

\section*{Acknowledgements}
The authors A. Barletta and M. Celli acknowledge the financial support from the Italian Ministry of Education, University and Research (MIUR), grant PRIN~2017F7KZWS.


\numberwithin{equation}{section}

\begin{appendices}

\section{--~~~Batchelor's profile}\label{Appe}
If one considers the rectangular cavity displayed in Fig.~\ref{fig1} and assumes $H \gg L$, namely an infinitely tall cavity, then the velocity distribution at the midplane $y=0$ can be expressed as the cubic Batchelor's profile \cite{batchelor1954heat},
\eqn{
v = - \frac{\rho_0\, \beta\, g\, \qty(T_1 - T_2)}{12\, \mu\, L}\, x\, \qty(L^2 - x^2) ,
}{a1}
with the linear temperature profile,
\eqn{
T - T_0 = - \frac{T_1 - T_2}{2\,L}\, x .
}{a2}
Equation~(\ref{a2}) defines the pure conduction regime happening in the cavity. 
On the basis of \equa{10}, \equasa{a1}{a2} yield the mass flow rate per unit depth (in the $z$ direction),
\eqn{
\dot{m} =  - \rho_0\, \beta \int\limits_{-L}^L \qty( T - T_0)\, v \, \dd x = - \frac{\rho_0^2\, \beta^2\, g\, \qty(T_1 - T_2)^2\, L^3}{90\,\mu} .
}{a3}

\end{appendices}

\end{document}